\documentclass[%
aip,
jcp,%
amsmath,amssymb,
reprint%
]{revtex4-1}
\usepackage[latin1]{inputenc}

\usepackage{amsmath}
\usepackage{amsfonts}
\usepackage{amssymb,bm}
\usepackage{graphicx}
\usepackage{physics}
\usepackage{upgreek}
\usepackage[outercaption]{sidecap}   
\usepackage{color}
\usepackage{txfonts}

\DeclareMathAlphabet{\pazocal}{OMS}{zplm}{m}{n}

\newcommand{\sing}{\ensuremath{\mathrm{S}}}
\newcommand{\trip}{\ensuremath{\mathrm{T}}}

\newcommand{\wig}{\ensuremath{\mathrm{W}}}

\renewcommand{\op}[1]{\ensuremath{\hat{#1}}}

\newcommand{\sbX}{\ensuremath{\boldsymbol{\mathsf{X}}}}
\newcommand{\CPFrp}{$\text{C}^{\bullet +}\text{PF}^{\bullet-}$}

\begin{document}
	
\title{Quantum mechanical spin dynamics of a molecular magnetoreceptor}
\author{Lachlan P. Lindoy}
\affiliation{Department of Chemistry, University of Oxford, Physical and Theoretical Chemistry Laboratory, South Parks Road, Oxford, OX1 3QZ, UK}
\author{Thomas P. Fay}
\affiliation{Department of Chemistry, University of Oxford, Physical and Theoretical Chemistry Laboratory, South Parks Road, Oxford, OX1 3QZ, UK}
\author{David E. Manolopoulos}
\affiliation{Department of Chemistry, University of Oxford, Physical and Theoretical Chemistry Laboratory, South Parks Road, Oxford, OX1 3QZ, UK}

\begin{abstract}
Radical pair recombination reactions are known to be sensitive to extremely weak magnetic fields, and can therefore be said to function as molecular magnetoreceptors. The classic example is a carotenoid-porphyrin-fullerene (\CPFrp) radical pair that has been shown to provide a \lq\lq proof-of-principle" for the operation of a chemical compass [K. Maeda {\em et al.}, Nature {\bf 453}, 387 (2008)]. Previous simulations of this radical pair have employed semiclassical approximations, which are routinely applicable to its 47 coupled electronic and nuclear spins. However, calculating the exact quantum mechanical spin dynamics presents a significant challenge, and has not been possible before now. Here we use a recently developed method to perform numerically converged simulations of the \CPFrp quantum mechanical spin dynamics, including all coupled spins. Comparison of these quantum mechanical simulations with various semiclassical approximations reveals that, while it is not perfect, the best semiclassical approximation does capture essentially all of the relevant physics in this problem. 
\end{abstract}

\maketitle
	
\section{Introduction}

\begin{figure}[t]
\includegraphics[width=0.49\textwidth]{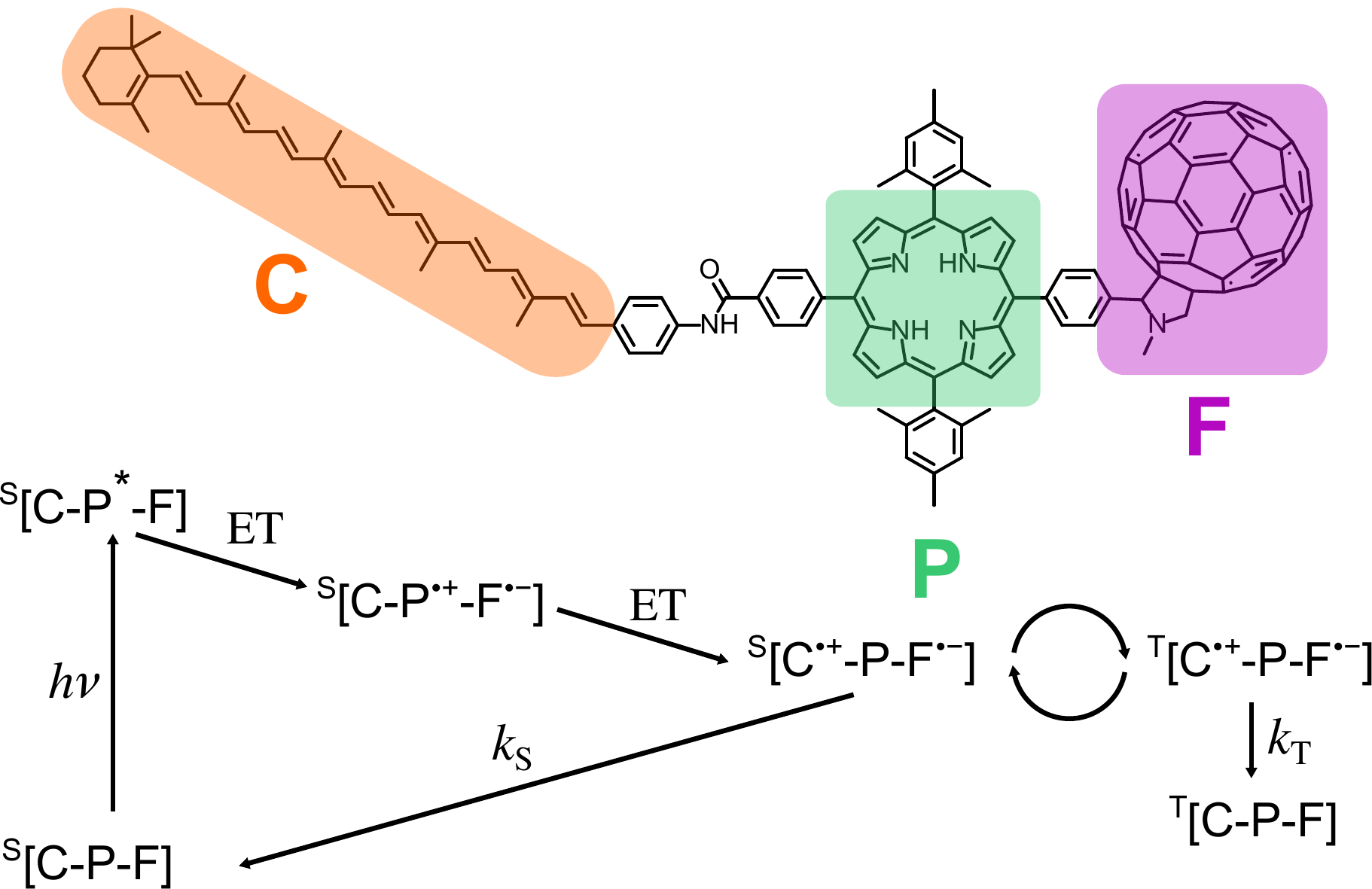}
\caption{The chemical structure and photophysics of the CPF triad molecule.}\label{fig1}
\end{figure}

The possibility that the quantum dynamics of the spins in radical pairs could underlie the magnetic compass sense of migratory songbirds has attracted a great deal of recent interest.\cite{Ritz2000,Maeda2012,Engels2014,Lee2014,Holland2014,Kattnig2016,Hiscock2016,Hore2016,Nordmann2017,Gunther2018,Mouritsen2018,Hiscock2019,Wiltschko2019} One indication that this might be possible is provided by the first experimental demonstration that a radical pair recombination reaction can function as a magnetoreceptor in an Earth strength magnetic field.\cite{Maeda2008,Maeda2011,Kerpal2019} A carotenoid-porphyrin-fullerene (CPF) molecule rapidly undergoes two successive electron transfer reactions after photoexcitation to form a long-lived \CPFrp radical pair, predominantly in its electronic singlet state, as illustrated in Fig. \ref{fig1}. Once the radical pair has formed, hyperfine interactions between the electron and nuclear spins in the carotenoid radical cause coherent transitions between the singlet and triplet radical pair states, a process which is also affected by the Zeeman interaction of the electron spins with an applied magnetic field.  Because the singlet and triplet radical pair states decay at different rates, the time-dependent survival probability of the carotenoid C$^{\bullet +}$ radical in the pair, which is detectable by transient absorption spectroscopy,\cite{Maeda2008,Maeda2011,Kerpal2019} is also sensitive to the applied magnetic field. This sensitivity has been detected experimentally in magnetic fields as low as 39 $\mu$T,\cite{Maeda2008} and also simulated theoretically with semiclassical spin dynamics calculations.\cite{Lewis2014,Lewis2018} 

Since the carotenoid radical contains 45 protons with significant hyperfine coupling constants, going beyond the semiclassical approximation to the spin dynamics presents a considerable challenge. The full spin system has a Hilbert space of dimension $2^{47} > 10^{14}$, and a Liouville space of dimension $2^{94} > 10^{28}$, which makes a naive brute force calculation of the quantum mechanical spin dynamics quite impractical. Using efficient sampling techniques, the upper limit that such a calculation can currently reach is a system with $\sim \!20$ coupled nuclear spins,\cite{Lewis2016,Fay2017} which is far fewer than are present in \CPFrp. A number of semiclassical methods have been developed with the aim of simulating the spin dynamics of radical pairs as complex as this,\cite{Lewis2014,Schulten1978,Manolopoulos2013,Fay2020} but until recently it has not been possible to validate these methods by comparison with exact quantum calculations for such large spin systems.

Here we use a recently developed method based on systematically approximating the radical pair Hamiltonian,\cite{Lindoy2018} and efficiently sampling the resulting quantum mechanical traces,\cite{Lewis2016} to perform numerically converged quantum dynamics simulations of the \CPFrp radical pair. This is the the first time that a fully quantum mechanical calculation of radical pair spin dynamics has been performed for a system with so many coupled nuclear spins. Comparison of the converged quantum mechanical results with semiclassical simulations enables us to assess the accuracy of various semiclassical approximations to the spin dynamics.

In Sec.~\ref{theory-sec} we outline the methods we have used to perform simulations of the \CPFrp radical pair. We describe the techniques that have enabled us to perform numerically exact quantum mechanical calculations, as well as the semiclassical methods we have used to approximate the spin dynamics. In Sec.~\ref{cpf-sim-sec} we describe the model parameters we have used for \CPFrp and provide some details of the simulations, including a demonstration of the convergence of our quantum mechanical calculations to graphical accuracy. The results of the quantum and semiclassical spin dynamics simulations are presented in Sec.~\ref{results-sec}, and our conclusions are drawn in Sec.~\ref{conc-sec}. 

\section{Theory}\label{theory-sec}

\subsection{Spin dynamics of radical pairs}

The spin degrees of freedom of a radical pair are described by its spin density operator $\op{\rho}(t)$, which satisfies the Haberkorn master equation\cite{Haberkorn1976,Ivanov2010,Fay2018}
\begin{align}\label{haberkorn-qme-eq}
	\dv{t}\op{\rho}(t) = -i \left[\op{H},\op{\rho}(t)\right]_- - \left[\op{K},\op{\rho}(t)\right]_+. 
\end{align}
Here $\op{H}$ is the spin Hamiltonian, $\op{K}$ is the Haberkorn reaction operator, $[\op{A},\op{B}]_\pm = \op{A}\op{B}\pm\op{B}\op{A}$, and we have set $\hbar=1$ (as we shall do throughout the following). The Haberkorn reaction operator is 
\begin{align}
	\op{K} = \frac{k_\sing}{2}\op{P}_\sing + \frac{k_\trip}{2}\op{P}_\trip,
\end{align}
where $k_\sing$ and $k_\trip$  are the total first-order singlet and triplet recombination rate constants and $\op{P}_\sing$ and $\op{P}_\trip$ are projection operators onto the electronic singlet and triplet subspaces of the radical pair. These can be written in terms of the electron spin operators $\op{\vb{S}}_1$ and $\op{\vb{S}}_2$ as
\begin{subequations}
	\begin{align}
		\op{P}_\sing &= \frac{1}{4}\op{1} - \op{\vb{S}}_1\cdot \op{\vb{S}}_2 \\
		\op{P}_\trip &= \frac{3}{4}\op{1} + \op{\vb{S}}_1\cdot \op{\vb{S}}_2.
	\end{align}
\end{subequations}

The spin Hamiltonian, in which we will only consider the isotropic hyperfine coupling and isotropic Zeeman terms, can be written as a sum of single radical Hamiltonians $\op{H}_i$,\cite{Steiner1989}
\begin{align}
	\op{H} = \op{H}_1 + \op{H}_2.
\end{align}
Each single radical Hamiltonian contains a Zeeman interaction for the electron spin and a set of isotropic hyperfine interactions,\cite{Steiner1989}
\begin{align}
	\op{H}_i = \boldsymbol{\omega}_i \cdot \op{\vb{S}}_i + \sum_{k=1}^{N_i}a_{ik}\op{\vb{I}}_{ik}\cdot\op{\vb{S}}_{i}.
\end{align}
Here $\boldsymbol{\omega}_i = g_i \mu_\mathrm{B} \vb{B}$ is the Zeeman frequency of the electron spin, which depends on its $g$-value $g_i$ and the applied magnetic field $\vb{B}$. $\op{\vb{I}}_{ik}$ is the vector operator of a nuclear spin with spin angular momentum quantum number $I_{ik}$, and $a_{ik}$ is the hyperfine coupling constant for this nuclear spin. 

We can usually assume there are no correlations between electron and nuclear spins at $t=0$ and that the nuclear spins are initially in a completely mixed state, so the initial spin density operator can be written as
\begin{align}
	\op{\rho}(0) = \frac{1}{Z}\op{\sigma}(0),
\end{align}
where $Z$ is the dimensionality of the nuclear spin Hilbert space and $\op{\sigma}(0)$ is a normalised electron spin density operator. For example, for a singlet-born radical pair, $\op{\sigma}(0) = \op{P}_\sing$. The expectation value of an observable $O$ of the spin system at time $t$ is then given by 
\begin{align}
  \ev{O(t)} = \Tr[\op{O}\op{\rho}(t)].
\end{align}
Solving Eq. \eqref{haberkorn-qme-eq}, observables can be written in terms of correlation functions of the form
\begin{align}\label{corr-func-eq}
  \ev{A(0)B(t)} = \Tr[\op{A}e^{+i\op{H}t-\op{K}t}\op{B}e^{-i\op{H}t-\op{K}t}].
\end{align}
For example, the time dependent singlet and triplet radical pair survival probabilities of a singlet-born radical pair are given by
\begin{subequations}
\begin{align}
  p_\sing(t) &= \frac{1}{Z}\ev{P_\sing(0)P_\sing(t)}\\
  p_\trip(t) &= \frac{1}{Z}\ev{P_\sing(0)P_\trip(t)}.
\end{align}
\end{subequations}
If there exists a basis in which $\op{P}_\sing$, $\op{P}_\trip$, and $\op{H}$ all have real matrix representations, as is the case for the Hamiltonian in Eqs.~(4) and~(5), then the correlation function $\ev{P_\sing(0)P_\trip(t)}$ is equivalent quantum mechanically to $\ev{P_\trip(0)P_\sing(t)}$ (see appendix~A). Since the semiclassical methods we shall employ do not all satisfy this exact symmetry constraint, there is some freedom in how to calculate time-dependent observables with these methods, which we shall explore.

\subsection{Quantum dynamics}

In order to perform numerically converged quantum dynamical calculations of a radical pair with as many as 47 coupled spins, we employ a method based on fitting a sequence of approximate, high symmetry, Hamiltonians to the Hamiltonian of each radical in the pair.\cite{Lindoy2018} In the \CPFrp radical pair, all $I_{ik}=1/2$, so we shall restrict our discussion to this case. As in Ref. \onlinecite{Lindoy2018}, the $M_i$-th approximation to the Hamiltonian of radical $i$ is written as
\begin{align}
\op{H}_i^{(M_i)} = \boldsymbol{\omega}_i \cdot \op{\vb{S}}_i + \sum_{j=1}^{M_i} \tilde{a}_{ij} \sum_{k=1}^{N_{ij}}\op{\vb{I}}_{ijk}\cdot \op{\vb{S}}_{i}
\end{align}
where $\op{\vb{I}}_{ijk}$ is a spin 1/2 nuclear spin operator and the parameters $\tilde{a}_{ij}$ and $N_{ij}$ are chosen so that the first $M_i+1$ moments of the approximate hyperfine distribution coincide with those of the exact hyperfine distribution; i.e., such that
\begin{align}
\mu_n^{(i)} = \sum_{k=1}^{N_i} a_{ik}^n = \sum_{j=1}^{M_i} N_{ij}\tilde{a}_{ij}^n
\end{align}
for $i=0,\ldots,M_i$. 

The dynamics generated by $\hat{H}_{i}^{(M_i)}$ can be calculated much more efficiently than that generated by the original Hamiltonian $\hat{H}_i$ when $M_i\ll N_i$, because the symmetry of $\hat{H}_{i}^{(M_i)}$ can then be used to separate the overall calculation into a set of much cheaper calculations in smaller Hilbert sub-spaces. Moreover $M_i$ can be systematically increased until the spin dynamics is found to converge, which typically happens in practice for $M_i \ll N_i$. 

We use the following procedure to find the parameters $N_{ij}$ and $\tilde{a}_{ij}$ in Eq.~(11).\cite{Lindoy2018} First we use a discrete procedure of Stieltjes\cite{Press1992} to construct a Gaussian quadrature rule with non-integer weights $N^{(0)}_{ij}$ and nodes $\tilde{a}^{(0)}_{ij}$ that captures the first $2M_{i}+1$ moments of the hyperfine distribution. The weights of this quadrature rule are then each either rounded up ($N_{ij}=\lceil {N}^{(0)}_{ij}\rceil$) or down ($N_{ij}=\lfloor {N}^{(0)}_{ij}\rfloor$) to give a set of integer values of $N_{ij}$. For each such set satisfying $\sum_{j=1}^{M_i}N_{ij}=N_i$, the nodes $\tilde{a}^{(0)}_{ij}$ are used as a starting point for solving the $M_i+1$ moment equations for the $\tilde{a}_{ij}$ in Eq.~(11), using Newton's method. Finally, we choose the set of $N_{ij}$ and $\tilde{a}_{ij}$ from among the resulting solutions that minimises the error in $\mu_{M_i+1}^{(i)}$. This procedure may seem complicated, but it is actually quite straightforward to implement. A computer program that implements it is provided in the supplementary material of Ref.~\onlinecite{Lindoy2018}.

The approximate Hamiltonian in Eq.~(10) contains $M_i$ sets of equivalent spin 1/2 nuclei, and it is this symmetry that is exploited to accelerate the calculation of the radical pair spin dynamics. $\op{H}_i^{(M_i)}$ commutes with $\op{K}_{ij}^2 = \op{\vb{K}}_{ij}\cdot\op{\vb{K}}_{ij}$, where $\op{\vb{K}}_{ij} = \sum_{k=1}^{N_{ij}} \op{\vb{I}}_{ijk}$, and therefore the Hamiltonian block diagonalises into subspaces of states which are eigenstates of $\op{K}_{ij}^2$. The eigenvalues of $\op{K}_{ij}^2$ in these blocks are $K_{ij}(K_{ij}+1)$, where $K_{ij}=(N_{ij}/2 - \lfloor N_{ij}/2\rfloor),\ldots ,N_{ij}/2-1,N_{ij}/2 $. Furthermore, for a given value of $K_{ij}$, there are multiple identical blocks. The number of these blocks, $w_{ij}(K_{ij})$, is 
\begin{align}
  w_{ij}(K_{ij}) = {{N_{ij}}\choose {N_{ij}/2}+K_{ij}}{2K_{ij}+1\over N_{ij}/2+K_{ij}+1}.
\end{align}

Because the initial density operator and the electron spin observables also commute with $\op{K}_{ij}^2$, the expectation values of these observables can be written as
\begin{align}
  \ev{O(t)} = \sum_{\vb{K}} w_{\vb{K}} \Tr_{\vb{K}} {[\op{O}\op{\rho}(t)]},
\end{align}
where the $\Tr_{\vb{K}}$ denotes a trace restricted to the symmetry block with total angular momentum quantum numbers $\vb{K} = K_{1,1}, K_{1,2},\dots,K_{2,M_2}$, and $w_{\vb{K}}$ is given by
\begin{align}
  w_{\vb{K}} = \prod_{i=1}^2\prod_{j=1}^{M_i}w_{ij}(K_{ij}).
\end{align}

For a singlet born radical pair with $\op{\rho}(0) = \op{P}_\sing/Z$, the individual block calculations can be further reduced to a set of independent wave function evolutions using the fact that
\begin{align}
  w_{\vb{K}}\Tr_{\vb{K}} {[\op{O}\op{\rho}(t)]} = \frac{w_{\vb{K}}}{Z}\sum_{\vb{M}}\ev{\op{O}}{\sing,\vb{K},\vb{M};t}
\end{align}
where 
\begin{align}\label{evolved-state-eq}
  \ket{\sing,\vb{K},\vb{M};t} = \exp(-i\op{H}t-\op{K}t)\ket{\sing,\vb{K},\vb{M};0},
\end{align}
with
\begin{align}
  \ket{\sing,\vb{K},\vb{M};0}=\ket{\sing}\otimes\left(\bigotimes_{i=1}^2\bigotimes_{j=1}^{M_i}\ket{K_{ij},M_{ij}}\right).
  \end{align}

The trace in Eq.~(15) can still be very expensive to evaluate when there are a large number of nuclear spin states in the symmetry block. This bottleneck can be overcome by using coherent spin state sampling to evaluate the trace.\cite{Lewis2016} The trace can be re-written exactly in terms of an integral over coherent spin states as\cite{Radcliffe1971,Arecchi1972}
\begin{align}
  w_{\vb{K}}\Tr_{\vb{K}} {[\op{O}\op{\rho}(t)]} = \frac{w_{\vb{K}}}{Z}\int\dd{\boldsymbol{\Omega}}\ev{\op{O}}{\sing,\vb{K},\boldsymbol{\Omega};t}
\end{align}
where $\ket{\sing,\vb{K},\boldsymbol{\Omega};t}$ is a time evolved state as in Eq. \eqref{evolved-state-eq}, but initialised in an electronic singlet state and a nuclear spin coherent state, $\ket{\sing,\vb{K},\boldsymbol{\Omega};0}=\ket{\sing}\otimes\left(\bigotimes_{i=1}^2\bigotimes_{j=1}^{M_i}\ket{K_{ij},\Omega_{ij}}\right)$,\cite{Lewis2016,Fay2017} where $\ket{K_{ij},\Omega_{ij}}$ is the $\ket{K_{ij},K_{ij}}$ state with the axis of quantisation rotated to lie in the direction $\Omega_{ij}=(\theta_{ij},\phi_{ij})$.\cite{Radcliffe1971,Arecchi1972} The nuclear spin coherent states are thus parameterised by these angles, and we integrate each set of angles over the surface of a sphere,\cite{Radcliffe1971,Arecchi1972}
\begin{align}
  \int\dd{\boldsymbol{\Omega}} = \prod_{i=1}^2\prod_{j=1}^{N_{ij}} \frac{2K_{ij}+1}{4\pi}\int_0^{2\pi}\dd{\phi_{ij}}\int_{0}^{\pi}\sin{\theta_{ij}}\dd{\theta_{ij}}.
\end{align}
These integrals can be evaluated by Monte Carlo sampling,\cite{Lewis2016} which we do whenever the size of the nuclear spin subspace $Z_{\vb{K}} = \prod_{ij}(2K_{ij}+1)$ is large enough to make this more efficient than a deterministic evaluation of the trace using Eq.~(15). 

A final tweak is to note that we can simply discard the symmetry blocks for which $w_{\vb{K}}Z_{\vb{K}}/Z$ is below a predetermined threshold value, since these blocks will only make a negligible contribution to $\left<O(t)\right>$. In particular, because the operators $\hat{O}(t)=\hat{P}_{\rm S}(t)$ and $\hat{P}_{\rm T}(t)$ both have eigenvalues between 0 and 1, requiring that
\begin{align}
\sum_{{\bf K}\ {\rm discarded}} {w_{\bf K}Z_{\bf K}\over Z} < \epsilon
\end{align} 
is sufficient to ensure that the error in the computed $\left<O(t)\right>$ will be less than $\epsilon$. This results in a considerable computational saving, because the symmetry blocks with the smallest values of $w_{\bf K}Z_{\bf K}$ also have the largest values of $Z_{\bf K}$, and their traces are therefore the most expensive to evaluate using Eq.~(18).

\subsection{Semiclassical dynamics}

In addition to performing quantum dynamical calculations, we shall calculate the spin dynamics using two semiclassical methods.\cite{Schulten1978, Lewis2014} In both of these methods, the nuclear spin operators are mapped onto classical variables $\op{\vb{I}}_{ik}\to\vb{I}_{ik}$, along with the electron spin operators $\op{\vb{S}}_i \to \vb{S}_i$, the two-electron spin operators $\op{S}_{1\alpha}\op{S}_{2\beta}=\op{T}_{\alpha\beta} \to T_{\alpha\beta}$, and the identity operator $\op{1}\to\bar{1}$. In the following we will use $\vb{I}$ to denote the complete set of classical nuclear spin variables $\vb{I} = \vb{I}_{1,1},\ldots,\vb{I}_{2,N_2}$, and $\sbX$ to denote the set of classical variables for the one- and two-electron spin operators and the identity operator. 

The electron spin correlation functions that we are interested in can be approximated as averages over independent trajectories of these classical variables as
\begin{align}
  \ev{A(0)B(t)}\!\approx\!\!\int\!\!\!\dd{\sbX}\!\!\!\int\!\!\!\dd{\vb{I}} \mu(\sbX,\vb{I}) A^\wig\!(\sbX,\vb{I}) B^\wig\!(\sbX(t),\vb{I}(t)),
\end{align}
where $A^\wig(\sbX,\vb{I})$ and $B^\wig(\sbX,\vb{I})$ are the phase space representations of $\op{A}$ and $\op{B}$ constructed using the above mapping. For example, the expression for $\hat{P}_{\rm S}$ in Eq.~(3a) gives $P_\sing^\wig(\sbX,\vb{I}) = (\bar{1}/4-\sum_{\alpha}T_{\alpha\alpha})$. The phase space measure $\mu(\sbX,\vb{I})$ is given by
\begin{align}
  \mu(\sbX,\vb{I}) = \mu_{12}(\sbX) \prod_{i=1}^2\prod_{k=1}^{N_i}\mu_{I_{ik}}(\vb{I}_{ik}).
\end{align}
in which $\mu_{12}(\sbX)$ is 
\begin{align}
\begin{split}
\mu_{12}(\sbX) =  &\frac{4}{(3\pi)^2 }\,{\delta(|\vb{S}_1|-\sqrt{3}/2)}\,{\delta(|\vb{S}_2|-\sqrt{3}/2)} \\
&\times \delta({\bar{1}-1})\prod_{\alpha\beta}\delta(T_{\alpha\beta} - S_{1\alpha}S_{2\beta}),
\end{split}
\end{align}
and $\mu_{I_{ik}}(\vb{I}_{ik})$ is
\begin{align}
\mu_{I_{ik}}(\vb{I}_{ik}) = \frac{2}{3\pi} \,{\delta(|\vb{I}_{ik}|-\sqrt{3}/2)}.
\end{align}
The time-dependent electronic and nuclear spin variables $\sbX(t)$ and $\vb{I}(t)$ are initially set to $\sbX(0)=\sbX$ and $\vb{I}(0) = \vb{I}$, and they evolve according to the following semiclassical equations of motion,\cite{Lewis2014}

\begin{subequations}
  \begin{align}
    \begin{split}
      \dv{t}S_{1\alpha}(t) &= \epsilon_{\alpha\beta\gamma}\left( \omega_{1\beta} +\sum_{k=1}^{N_1}a_{1k}I_{1k\beta}(t) \right)S_{1\gamma}(t) \\
      &-\bar{k}S_{1\alpha}(t) +\Delta k S_{2\alpha}(t)
    \end{split}\\
    \begin{split}
      \dv{t}S_{2\alpha}(t) &= \epsilon_{\alpha\beta\gamma}\left( \omega_{2\beta} +\sum_{k=1}^{N_2}a_{2k}I_{2k\beta}(t) \right)S_{2\gamma}(t) \\
      &-\bar{k}S_{2\alpha}(t) +\Delta k S_{1\alpha}(t)
    \end{split}\\
     \begin{split}
      \dv{t}T_{\alpha\beta}(t) &= \epsilon_{\alpha\gamma\delta}\left( \omega_{1\gamma} +\sum_{k=1}^{N_1}a_{1k}I_{2k\gamma}(t) \right)T_{\delta\beta}(t) \\
      &+\epsilon_{\beta\gamma\delta}\left( \omega_{2\gamma} +\sum_{k=1}^{N_2}a_{2k}I_{2k\gamma}(t) \right)T_{\alpha\delta}(t) \\
      &-\bar{k}T_{\alpha\beta}(t)-\Delta k T_{\beta\alpha}(t) + \delta_{\alpha\beta}\Delta k\left(\,\frac{\bar{1}(t)}{4} \!-\! T_{\gamma\gamma}(t)\,\right)
    \end{split}\\
    \dv{t}\bar{1}(t)\phantom{x} &= -\bar{k}\bar{1}(t)+ 4\Delta k\,T_{\alpha\alpha}(t) \\
    \dv{t}I_{ik\alpha}(t) &= a_{ik} \epsilon_{\alpha\beta\gamma}\left(\frac{\sqrt{3}}{2}\frac{S_{i\beta}(t)}{|{\bf S}_i(t)|}\right)I_{ik\gamma}(t). \label{nuc-eom-eqn}
  \end{align}
\end{subequations}
Here $\bar{k} = (k_\sing+3k_\trip)/4$, $\Delta k = (k_\sing-k_\trip)/4$, $\epsilon_{\alpha\beta\gamma}$ is the alternating tensor, $\delta_{\alpha\beta}$ is the Kronecker delta, and we have used the summation convention for repeated Greek (cartesian coordinate) indices. 

When $k_\sing\neq k_\trip$, the semiclassical approximations to $\ev{P_\sing(0) P_\trip(t)}$ and $\ev{P_\trip(0) P_\sing(t)}$ are not equivalent, which gives us two different options for evaluating $p_\trip(t)$. In the following we shall consider calculating $p_\trip(t)$ using both $\ev{P_\sing(0) P_\trip(t)}$ and $\ev{P_\trip(0) P_\sing(t)}$, which we shall refer to as the SC (a) and SC (b) methods respectively. 

As well as performing semiclassical spin dynamics calculations using the method outlined above, we shall also use the Schulten-Wolynes (SW) method.\cite{Schulten1978} In the present context, this method can be obtained by simply setting the right-hand side of the equation of motion for the nuclear spin variables, Eq.~\eqref{nuc-eom-eqn}, to zero.  (In the formulation originally presented by Schulten and Wolynes,\cite{Schulten1978} the central limit theorem was invoked to approximate the distribution of the overall hyperfine field in each radical as a Gaussian. For the model \CPFrp radical pair that we shall study, this additional approximation is almost certainly justified, but we have not actually made it in our calculations.) Note that when the evolution of the nuclear spin variables in Eq.~(25e) is suppressed, the semiclassical $\ev{P_\sing(0) P_\trip(t)}$ and $\ev{P_\trip(0) P_\sing(t)}$ correlation functions become equivalent, so unlike in the above semiclassical method, there is only one way to evaluate $p_\trip(t)$ in the SW method.

\begin{figure}[t]
\includegraphics[width=0.42\textwidth]{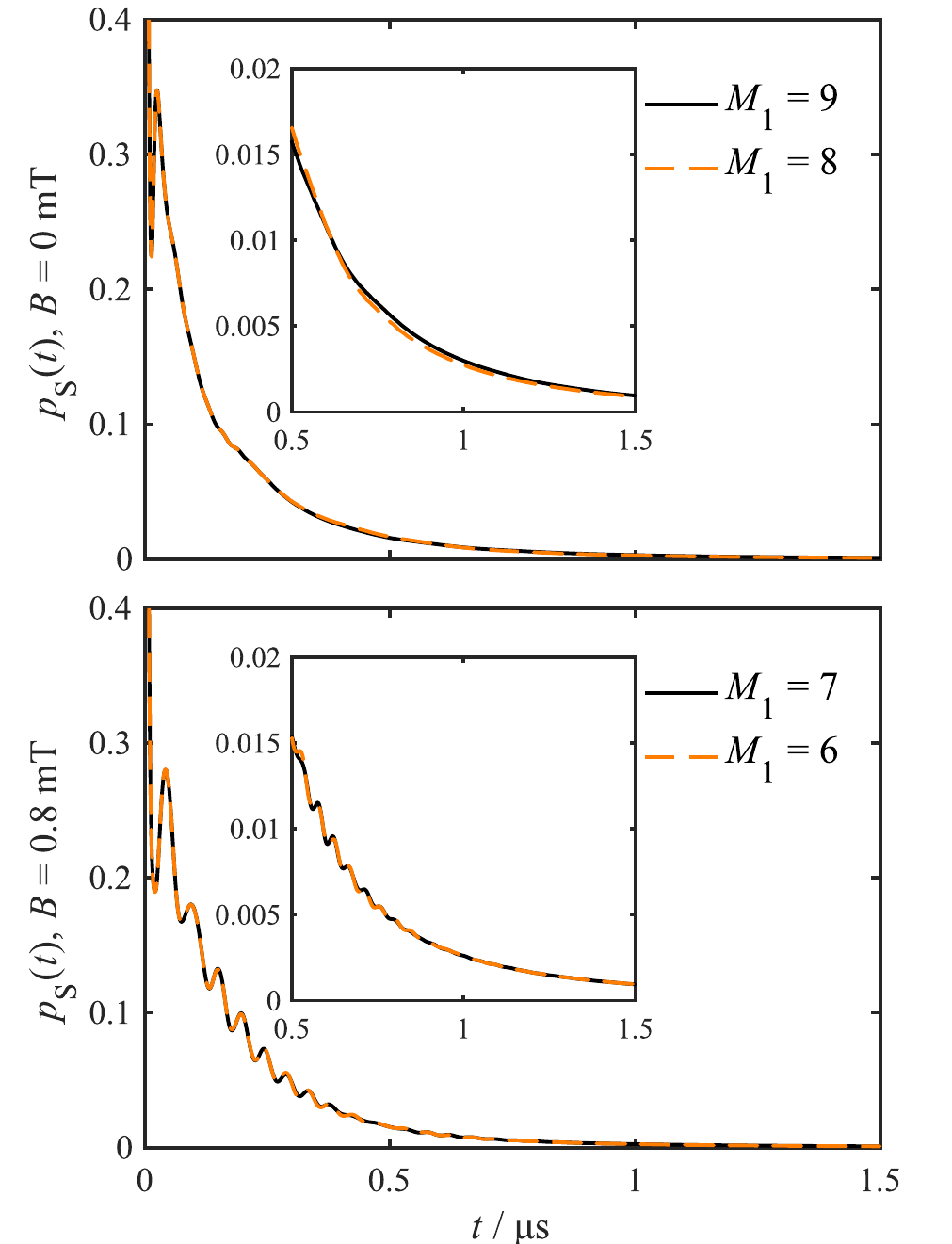}
\caption{Convergence of the quantum mechanical singlet probability of the \CPFrp radical pair with respect to $M_1$, for $B=0$ mT (top) and $B=0.8$ mT (bottom). Since we are neglecting $^{13}$C nuclei, the fullerene radical does not have any nuclear spins, so there is no need to demonstrate convergence with respect to $M_2$.}\label{fig2}
\end{figure}

\begin{figure*}[t]
\begin{minipage}[c]{0.75\textwidth}
\includegraphics[width=\textwidth]{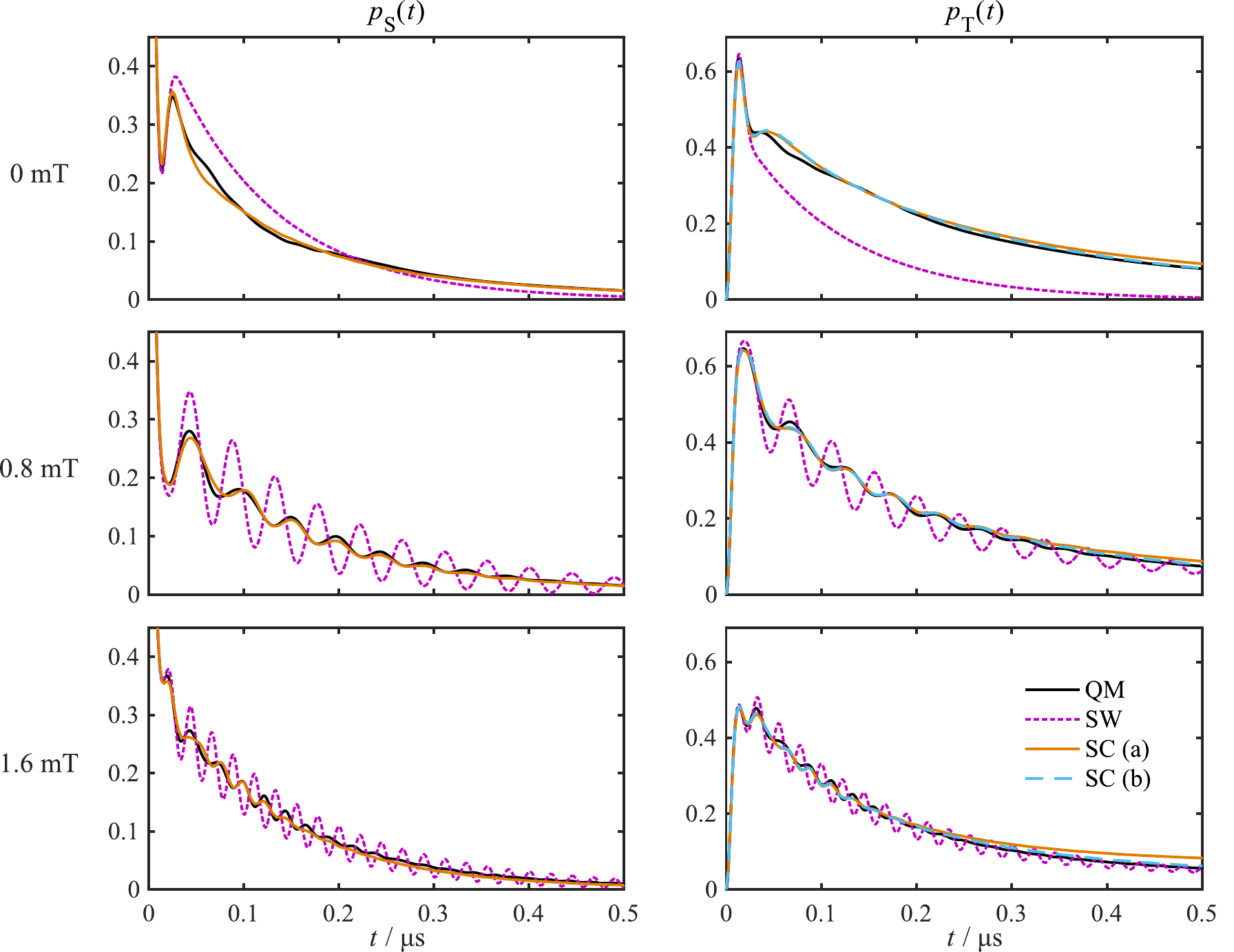}
\end{minipage}
\hfill
\begin{minipage}[c]{0.24\textwidth}
\caption{Singlet and triplet survival probabilities of the \CPFrp radical pair calculated quantum mechanically (QM), semiclassically (SC), and with the Schulten-Wolynes (SW) method. The SC (a) method uses the $\ev{P_\sing(0)P_\trip(t)}$ correlation function for $p_\trip(t)$, and SC (b) method uses $\ev{P_\trip(0)P_\sing(t)}$. Since both methods use $\ev{P_\sing(0)P_\sing(t)}$ for $p_\sing(t)$, they give the same result for the singlet probability, and only the SC (a) curve is shown.}\label{fig3}
\end{minipage}
\end{figure*}

\section{Simulation details}\label{cpf-sim-sec}

In our model for the \CPFrp radical pair we assume that both the carotenoid and fullerene radicals have isotropic $\vb{g}$ tensors with $g$-values equivalent to that of the free electron, $g_i = g_\mathrm{e}$. This approximation is valid since we only consider magnetic field strengths up to 1.6 mT, so the $\Delta g$ mechanism will not play an important role in the spin dynamics. We also assume that the scalar coupling between the two electrons in the radical pair is small, so we neglect this along with the anisotropic dipolar coupling. The isotropic hyperfine coupling constants of the carotenoid protons are listed in Appendix \ref{hfc-appendix}, and we ignore the presence of any ${}^{13}\text{C}$ nuclei in the radical pair. We take the singlet and triplet first order recombination rate constants to be $k_\sing = 1.8\times 10^{7}\text{ s}^{-1}$ and $k_\trip = 7.1\times 10^{4}\text{ s}^{-1}$, as estimated from EPR experiments on the radical pair in solution at 110 K.\cite{Maeda2011} In order to perform the quantum mechanical calculations on the 47 spin system, we have to neglect the effect of electron spin relaxation, so we also neglect this in the semiclassical calculations. The \CPFrp radical pair is formed primarily in the singlet state after photoexcitation of CPF at 110 K,\cite{Maeda2011} so we ignore the presence of any initial triplet radical pairs.

In our quantum mechanical calculations, we used Eq.~(15) to evaluate the traces of symmetry blocks with $Z_{\bf K}\le 500$. The traces of the larger symmetry blocks were evaluated using Eq.~(18), with 500 Monte Carlo samples of the initial nuclear spin coherent states.\cite{Lewis2016} The symmetry blocks with $w_{\bf K}Z_{\bf K}/Z<10^{-5}$ were deemed to make a negligible contribution to $p_{\rm S}(t)$ and $p_{\rm T}(t)$ and discarded. The short iterative Arnoldi method, a Krylov subspace method similar to the short iterative Lanczos method\cite{Park1986} but applicable to systems with non-unitary dynamics, was used to propagate the spin states forwards in time.

Since the fullerene radical does not contain any hyperfine coupled nuclei in our model, we only needed to fit the carotenoid radical Hamiltonian $\hat{H}_1$ to a sequence of symmetrized Hamiltonians $\op{H}^{(M_1)}_1$. We explored various values of $M_1$ up to $M_1=10$, and found that the quantum mechanical results were well converged with $M_1=9$ for $B\le 0.4$ mT, and with $M_1=7$ for $B\ge 0.4$ mT.  Fig.~\ref{fig2} shows that these values of $M_1$ are sufficient to converge the singlet radical pair survival probabilities to within $10^{-3}$ over the time-scale of interest. The convergence of the triplet survival probabilities was found to be the much the same. In our semiclassical and Schulten-Wolynes calculations, we used a million Monte-Carlo samples to evaluate the integrals in Eq.~(21).

\section{Results and discussion}\label{results-sec}

\begin{figure*}[t]
\begin{minipage}[c]{0.75\textwidth}
\includegraphics[width=\textwidth]{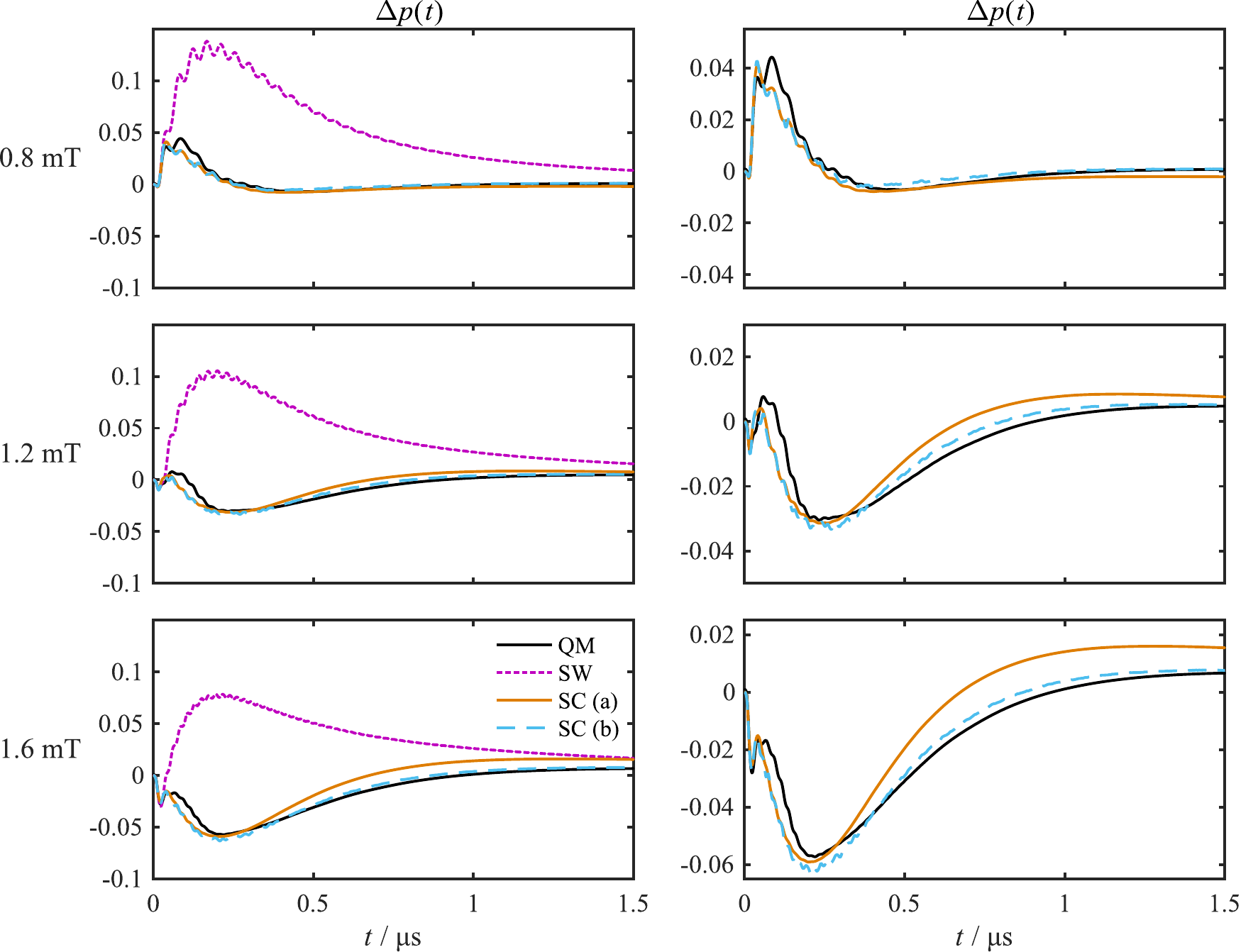}
\end{minipage}
\hfill
\begin{minipage}[c]{0.24\textwidth}
\caption{Magnetic field effect on the total radical pair survival probability $\Delta p(t,B)$. The left hand column shows the QM, SC, and  SW results for our model \CPFrp radical pair at $B=0.8$ mT, 1.2 mT, and 1.6 mT. The right hand column shows the QM and SC results on an expanded scale, so that they can be compared more easily.}\label{fig4}
\end{minipage}
\end{figure*}

Here we present the results of our spin dynamics calculations for the model \CPFrp radical pair defined in Sec.~III.

In Fig.~\ref{fig3} we show the time-dependent singlet and triplet radical pair populations at applied magnetic field strengths of 0 mT, 0.8 mT and 1.6 mT, calculated both quantum mechanically and semiclassically. The SW method is seen to agree well with the quantum simulation for very short times, up to about $20$ ns, but after that its accuracy degrades. The SC methods agree much better with the quantum simulation at longer times. For field strengths above 0 mT, a decaying Zeeman oscillation between the singlet and triplet states is introduced with a frequency of $\omega_{\rm e}=g_{\rm e}\mu_{\rm B}B$. This oscillation is captured qualitatively in all of the semiclassical calculations, but its amplitude is overestimated considerably by the SW method. At zero magnetic field, the SW method also significantly underestimates the degree of population transfer from the singlet state to triplet states. Both SC methods are reasonably accurate at zero field for the singlet populations, although they do miss a subtle quantum mechanical oscillation at around 50 ns. The SC (a) method consistently overestimates the triplet populations, which are captured significantly more accurately by the SC (b) method. 

We hypothesise that the SC (b) method performs better than the SC (a) method for the triplet populations because our model for the radical pair recombination has $k_\sing > k_\trip$. The observable that is propagated in the SC (b) method, $P_\sing^\wig(\sbX(t),\vb{I}(t))$, therefore decays more rapidly than the observable that is propagated in the SC (a) method, $P_\trip^\wig(\sbX(t),\vb{I}(t))$. Since the accuracy of the semiclassical approximation degrades with time, it is more accurate to propagate the more rapidly decaying of the two observables that are correlated in the semiclassical correlation function when both options would give the same result quantum mechanically.

In transient absorption experiments, the singlet and triplet survival probabilities are not directly accessible. Instead what is typically measured is the magnetic field effect on the total survival probability of the radical pair,
\begin{align}
\Delta p(t,B) = p(t,B) - p(t,0),
\end{align}
where $p(t,B) = p_\sing(t,B) + p_\trip(t,B)$ is the total survival probability at a given magnetic field strength $B$. This can be measured by detecting the transient absorption of the carotenoid radical, which is present in both the singlet and triplet states of the radical pair, at time $t$ after the initial photoexcitation laser pulse, as a function of $B$.\cite{Maeda2008,Lewis2018} Because we are ultimately interested in modelling these types of experiments, we have calculated $\Delta p(t,B)$ for the present \CPFrp model at magnetic field strengths of 0.8 mT, 1.2 mT, and 1.6 mT. The results of these calculations are shown in Fig.~4.

The quantum mechanical results in this figure display a biphasic to triphasic to inverted biphasic transition as the applied magnetic field strength is increased. A similar transition is observed experimentally,\cite{Lewis2018} and has been explained in terms of enhanced $\sing\leftrightarrow\trip_0$ interconversion at shorter times and lower fields, which decreases the overall decay rate because $k_\sing > k_\trip$, and diminished $\sing\leftrightarrow\trip_\pm$ interconversion at longer times and higher fields.\cite{Lewis2018} The SW method fails to even qualitatively capture this behaviour, as noted in Ref.~\onlinecite{Lewis2018}. This can largely be attributed to the fact that the SW approximation fails to capture the correct dynamics at zero field (see Fig.~\ref{fig3}). The SC methods do capture the biphasic to triphasic to inverted biphasic transition at least qualitatively, and in the case of the SC (b) method almost quantitatively. In particular, the SC (b) method is seen to be significantly more accurate than the SC (a) method at longer times. This reinforces our comments about the desirability of propagating the more rapidly decaying of the two correlated observables in the semiclassical time correlation function.

The SC (b) results in Fig.~4 are not perfect. The method does not precisely reproduce the quantum mechanical $\Delta p(t,B)$ signal. However, the shape of the signal is at least reproduced qualitatively, and it is certainly reproduced well enough to capture the correct physics of the biphasic-triphasic-inverted biphasic transition.\cite{Lewis2018} Since $\Delta p(t,B)$ is a subtle field-on minus field-off difference signal, which is more than an order of magnitude weaker than $p(t,B)$ itself (compare the ranges of the ordinates in Figs.~3 and~4), we find this to be very encouraging. In fact, we would even go so far as to argue on the basis of Fig.~4 that the semiclassical SC (b) method is accurate enough to capture essentially all of the relevant physics in this problem.

\section{Concluding remarks}\label{conc-sec}

In this paper, we have shown that the method presented in Ref.~\onlinecite{Lindoy2018} can be used to obtain numerically converged results for the quantum spin dynamics of a radical pair containing as many as 45 hyperfine-coupled nuclear spins. We have also used these results to assess the accuracy of various semiclassical approximations to the spin dynamics, and found that the most accurate of these [the SC (b) method in which $p_{\rm T}(t)$ is calculated from $\ev{P_\sing(0)P_\trip(t)}$] reproduces the quantum mechanical results extremely well (see Figs.~3 and~4).

The quantum mechanical results for our model of the \CPFrp radical pair do not agree perfectly with the available experimental data.\cite{Lewis2018} In our simulations, the transition from biphasic to triphasic to inverted biphasic behaviour of the radical pair survival probability occurs at higher magnetic fields than those observed experimentally.\cite{Lewis2018} This is most likely due to deficiencies in our model parameters, and in the physics included in the model itself. Our neglet of electron spin relaxation effects may be particularly important. These cannot be included at all easily in the present quantum mechanical method, but they are straightforward to include semiclassically.\cite{Lewis2014} One possible strategy for gaining further insight into the CPF experiments would therefore be to use an inexpensive method such as the Schulten-Wolynes method with a master equation approach to the spin relaxation,\cite{FayLindoy2019} or kinetic master equations,\cite{Steiner2018,Fay2019,Mims2019} to fit a model to experimental data at higher field strengths where these approximations are most reliable. This model could then be used to simulate the experimental data at lower field strengths using a more accurate semiclassical method, such as the SC (b) method we have benchmarked here. However, without this benchmarking against exact quantum calculations (albeit for a simplified model without any electron spin relaxation), it would be impossible to know just how reliable these semiclassical calculations would be.

\section*{Data Availability Statement}

The data that support the findings of this study are available within the article itself.

\acknowledgements

Lachlan Lindoy is supported by a Perkin Research Studentship from Magdalen College, Oxford, an Eleanor Sophia Wood Postgraduate Research Travelling Scholarship from the University of Sydney, a James Fairfax Oxford Australia Scholarship, and a grant from the Air Force Office of Scientific Research (Air Force Materiel Command, USAF award no. FA9550-14-1-0095). Thomas Fay is supported by a Clarendon Scholarship from Oxford University, an E.A. Haigh Scholarship from Corpus Christi College, Oxford, and by the EPRSC Centre for Doctoral Training in Theory and Modelling in the Chemical Sciences, EPSRC Grant No. EP/L015722/1. 

\appendix

\section{Symmetry of $\ev{P_\sing(0)P_\trip(t)}$}\label{corr-func-app}

Here we will show that for the quantum mechanical radical pair spin dynamics, $\ev{P_\sing(0)P_\trip(t)}=\ev{P_\trip(0)P_\sing(t)}$. First we note that for a correlation function of the form given in Eq. \eqref{corr-func-eq}, if $\op{A}$ and $\op{B}$ are hermitian, then $\ev{A(0)B(t)}$ is real:
\begin{align}
  \begin{split}
    \ev{A(0)B(t)}^* &= \Tr[\left(\op{A}e^{+i\op{H}t-\op{K}t}\op{B}e^{-i\op{H}t-\op{K}t}\right)^\dag] \\
    &=\Tr[e^{+i\op{H}^\dag t-\op{K}^\dag t}\op{B}^\dag e^{-i\op{H}^\dag t-\op{K}^\dag t}\op{A}^\dag] \\
    &= \Tr[\op{A}e^{+i\op{H}t-\op{K}t}\op{B}e^{-i\op{H}t-\op{K}t}] \\
    &= \ev{A(0)B(t)}.
  \end{split}
\end{align}
Now suppose there exists a basis in which $\op{A}, \op{B}, \op{H}$, and $\op{K}$ all have real matrix representations. Evaluating the trace in this basis we find
\begin{align}
  \begin{split}
    \ev{A(0)B(t)} &= \ev{A(0)B(t)}^* \\
  &=  \Tr[\left(\vb{A}e^{+i\vb{H}t-\vb{K}t}\vb{B}e^{-i\vb{H}t-\vb{K}t}\right)^*] \\
  &=  \Tr[\vb{A}^*e^{-i\vb{H}^*t-\vb{K}^*t}\vb{B}^*e^{+i\vb{H}^*t-\vb{K}^*t}] \\
  &=  \Tr[\vb{A}e^{-i\vb{H}t-\vb{K}t}\vb{B}e^{+i\vb{H}t-\vb{K}t}] \\
  &=  \Tr[\vb{B}e^{+i\vb{H}t-\vb{K}t}\vb{A}e^{-i\vb{H}t-\vb{K}t}] \\
  &= \ev{B(0)A(t)}.
  \end{split}
\end{align}
Because all the scalar coupling terms in the Hamiltonian in Eqs.~(4) and~(5) are rotationally invariant, we can choose $\boldsymbol{\omega}_1$ and $\boldsymbol{\omega}_2$ to lie in the $x,z$ plane. Then in the standard uncoupled spin basis the matrix representations of $\op{P}_\sing$, $\op{P}_\trip$, $\op{H}$ and $\op{K}$ will all be real, giving $\ev{P_\sing(0)P_\trip(t)}=\ev{P_\trip(0)P_\sing(t)}$.

Of course this symmetry is not guaranteed to hold when one makes an approximation to the spin dynamics, such as the semiclassical approximations we have discussed in Sec.~II. We have found that it does hold for these approximations in the case of symmetric recombination ($k_{\rm S}=k_{\rm T})$, but that the SC (a) and SC (b) methods give different results in the case of asymmetric recombination ($k_{\rm S}\not=k_{\rm T}$). We would also expect them to give different results in the case of symmetric recombination when an exchange coupling between the two electrons is included in the spin Hamiltonian. However, we have not yet investigated this in any detail because it is not relevant to the problem we have considered in the present paper.

\section{CPF hyperfine constants}\label{hfc-appendix}

In our \CPFrp radical pair model, we use the following set of hyperfine constants for the protons on the carotenoid radical cation. These are the same as those used in Ref. \onlinecite{Lewis2014} but with the methyl group proton hyperfine constants averaged, to reflect the fact that these groups rapidly rotate on the time-scale of the radical pair spin dynamics. The hyperfine constants $a_{1k}$, in mT, are: $0.048790$, $0.046328$, $-0.115098$, $-0.111317$, $-0.361254$, $0.130081$, $0.094903$, $-0.316911$, $0.094676$, $-0.021817$, $-0.140593$, $-0.087963$, $-0.071456$, $0.050581$, $-0.275215$, $0.056448$, $0.111917$, $-0.385563$, $0.329013$, $0.329013$, $0.329013$, $0.216954$, $0.216954$, $0.216954$, $0.170627$, $0.170627$, $0.170627$, $0.304986$, $0.304986$, $0.304986$, $0.173690$, $0.579152$, $0.057321$, $0.006161$, $-0.005099$, $-0.003271$, $0.018443$, $0.001563$, $-0.017735$, $0.014287$, $-0.028314$, $0.003183$, $0.238826$, $0.238826$, $0.238826$.

\end{document}